\documentstyle[prb,aps,epsfig,floats,goodfloat]{revtex}

\newcommand{\bc}{\begin{center}}
\newcommand{\ec}{\end{center}}
\newcommand{\be}{\begin{equation}}
\newcommand{\ee}{\end{equation}}
\newcommand{\beqn}{\begin{eqnarray}}
\newcommand{\eeqn}{\end{eqnarray}}
\begin{document}
\draft

\twocolumn[\hsize\textwidth\columnwidth\hsize\csname@twocolumnfalse%
\endcsname

\title{
Triviality of the Ground State Structure in Ising Spin Glasses
}

\author{Matteo Palassini and A. P. Young}
\address{Department of Physics, University of California, Santa Cruz, 
CA 95064}

\date{\today}

\maketitle

\begin{abstract}
We investigate the
ground state structure of the three-dimensional Ising spin glass
in zero field by determining how the ground state
changes in a fixed finite block far from the boundaries when the boundary
conditions are changed. We find that
the probability of a change in the block
ground state configuration 
tends to zero
as the system size
tends to infinity.
This indicates a trivial ground state
structure, as predicted by the droplet theory. Similar results are also
obtained in two dimensions. 

\end{abstract}

\pacs{PACS numbers: 75.50.Lk, 05.70.Jk, 75.40.Mg, 77.80.Bh}
]

Controversy remains over the nature of ordering in spin glasses below the
transition temperature, $T_c$, and two scenarios have been extensively
discussed. In the ``droplet model'' proposed by Fisher and
Huse\cite{fh} (see also Refs.~\onlinecite{bm,mcmillan,ns-old,ns-new}), the
structure of ``pure states'' is predicted to be trivial. This means that
there  is a
unique state\cite{zero-field} in the sense that 
correlations of the spins in a region far from the boundaries are independent
of the boundary conditions imposed. As a consequence, the order
parameter
distribution function\cite{parisi,mpv,by}, $P(q)$, is also trivial, i.e. is
a pair of delta functions at $q=\pm q_{EA}$ where $q_{EA}$ is the 
Edwards-Anderson order parameter.
In the alternative approach, one assumes that the basic structure of the
Parisi\cite{parisi,mpv,by} solution of the infinite range model applies
also to realistic short range systems. In this picture, $P(q)$ is a non-trivial
function because many thermodynamic states contribute to
the partition function, i.e. the pure state structure is non-trivial.
Monte Carlo simulations on short range models on small
lattices\cite{rby,marinari,zuliani,berg}, find a non-trivial $P(q)$ with a
weight at $q=0$ which is independent of system size (for the range of sizes
studied), as predicted by the Parisi theory.

Most numerical work
has concentrated on $P(q)$. By contrast, here we attempt to
determine the pure state structure by investigating whether spin
correlation functions
in a {\em finite} region\cite{ns-new} far from the boundary,
change when the boundary conditions are
changed.
It is interesting to investigate this question even at $T=0$, where
there are 
efficient algorithms for determining ground states, even though
$P(q)$ is 
trivial in this limit (for a continuous bond
distribution).
Here we show that the Ising spin glass in
{\em three dimensions}\/ which has a {\em finite 
transition temperature}\/\cite{ky,mari3d,hart,matteo} $T_c$,
has a trivial ground
state structure. We also find a trivial ground state
structure in the {\em two-dimensional}
Ising spin glass which has a transition at
{\em zero temperature} with long-range order at $T=0$.
Some of our results in two dimensions
have also been reported elsewhere\cite{py} (referred to as
PY). Similar results for two dimensions, as well as results for some
three-dimensional models (but none for a spin glass with a finite
$T_c$) have also been found
by Middleton\cite{midd}.

The Hamiltonian is given by
\begin{equation}
{\cal H} = -\sum_{\langle i,j \rangle} J_{ij} S_i S_j ,
\label{ham}
\end{equation}
where the sites $i$ lie on a 
simple cubic ($d=3$) or square lattice ($d=2$) with $N=L^d$ sites 
($L \le 10$ in $3d$, $L \le 30$ in $2d$) , $S_i=\pm
1$, and the $J_{ij}$ are nearest-neighbor interactions chosen according to a
Gaussian distribution with zero mean and standard deviation unity.
We determine the energy and spin configuration of the ground state for a given
set of bonds, initially
for periodic boundary conditions denoted by ``P''.
Next we impose anti-periodic conditions (``AP'') along {\em
one}\/ direction, which is equivalent to keeping periodic boundary conditions
and changing the sign of the
interactions along this boundary, and recompute the ground state. Finally we
change the sign of  half the bonds at random along this boundary, which we
denote by ``R''. 

To determine the ground state in three dimensions
we use a hybrid genetic algorithm introduced by Pal\cite{pal1,pal2}.
Starting from a population of
random configurations (``parents''), new configurations
(``offspring'') are generated by recombination (triadic crossover) and
mutation. The population is progressively reduced, 
with a bias towards the offspring with lower energy. 
The algorithm is hybridized with
a local optimization of the offspring (see Ref.\onlinecite{pal1} for details).
For each sample and boundary condition, we repeat the algorithm $n_r$ times,
see Table I,
and take the lowest energy state found.
Our values for the average
ground state energy are in agreement with those
of  Ref.~\onlinecite{pal2}.
For $L\leq 6$, we checked all our results
with a different method, which consists in repeating many times the
microcanonical simulated annealing 
algorithm introduced in Ref.\onlinecite{Ocampo}. 
This  algorithm was also used to prepare the initial populations
of the genetic algorithm for $L \ge 8$. We discuss later additional checks
that we performed for the largest sizes $L=8$ and 10.

\begin{table}[ht]
\begin{center}
\begin{tabular}{lrrrr}
L  &  $N_s$  & $n_r$ &   $ \langle E \rangle $\ \ \ \ \   &  
$ \langle E \rangle $ Ref.\onlinecite{pal2} \\
\hline
4  & 20000 & 3  & $ -106.59(4)$ &  $-106.609(9)$  \\
5  & 15000 & 3  & $ -210.26(7)$ &  $-210.22(3) $  \\
6  & 9450  & 3 &  $ -364.9(1)$  &  $-364.89(5) $  \\   
8  & 6646   & 2 & $ -868.1(2)$  &  $-868.1(2)  $  \\ 
10 & 3010   & 1 & $-1697.5(4)$  & $-1698.8(8)  $  \\ 
\end{tabular}
\end{center}
\caption{Parameters of the simulation and results for the ground state energy
in $d=3$. As a function of the size $L$ we show: 
number of samples $N_s$, number of runs per sample $n_r$, and the
average ground state energy $\langle E \rangle $, from our data and that of
Pal\cite{pal2}.
}
\end{table} 

In two dimensions, we used the Cologne spin glass server\cite{juenger},
which calculates {\em exact}
ground states of the Ising spin glass with periodic
boundary conditions.

In order to study the dependence of the spin configuration on
boundary conditions we consider a central block containing $N_B = L_B^d$
spins. We compute
the block spin overlap distribution
$P^B_{\alpha\beta}(q)$, where $\alpha$ and $\beta$ denote two boundary
conditions, $P, AP$ or $R$ here, and
\begin{equation}
P^B_{\alpha\beta}(q) = \left\langle \delta \left( q - q^B_{\alpha\beta}
\right) \right\rangle ,
\end{equation}
in which
\begin{equation}
q^B_{\alpha\beta} = {1\over N_B} \sum_{i=1}^{N_B} S_i^\alpha S_i^\beta 
\end{equation}
is the overlap between the block configurations with
$\alpha$ and $\beta$ boundary conditions, $S_i^\alpha$ is the value of $S_i$
in the ground state with the $\alpha$ boundary condition, 
and the brackets $\langle \cdots \rangle$ refer to an average over the
disorder. 

Since we work at $T=0$, each sample and pair $\alpha,\beta$
gives a single value for $q$. 
The self overlap distribution, $P^B_{\alpha\alpha}(q)$,
has weight only at $q = \pm 1$,
since the ground state is unique for a given boundary condition.
$P^B(q)$ is normalized
to unity i.e. $\int P^B(q)\, dq = 1$, it is symmetric,
and the allowed $q$-values are discrete with a separation of of
$\Delta q = 2 / N^B$, so
$P^B_{\alpha\alpha}(\pm 1) = {N^B / 4 }. $

If the configuration
in the block changes when the boundary conditions are changed
from $\alpha$ to $\beta$, 
then the block overlap, $q^B_{\alpha\beta}$,
will no longer be $\pm 1$. Hence 
$ 1 -  P^B_{\alpha\beta}(1) / P^B_{\alpha\alpha}(1) $
is the probability that the block ground state changes on changing the boundary
conditions. We will see that
\begin{equation}
1 -  { P^B_{\alpha\beta}(1)  \over P^B_{\alpha\alpha}(1) }
\sim L^{-\lambda} ,
\label{def_lambda}
\end{equation}
with $\lambda > 0$, 
showing that the block ground state
configuration is unchanged in the thermodynamic limit.
Our interpretation\cite{py,midd} of this result is that the boundary condition
change induces a domain wall of fractal
dimension $d_f = d - \lambda$, and $L^{-\lambda}$ is then the
probability that the
domain wall intersects the block.

\begin{figure}[ht]
\begin{center}
\epsfig{figure=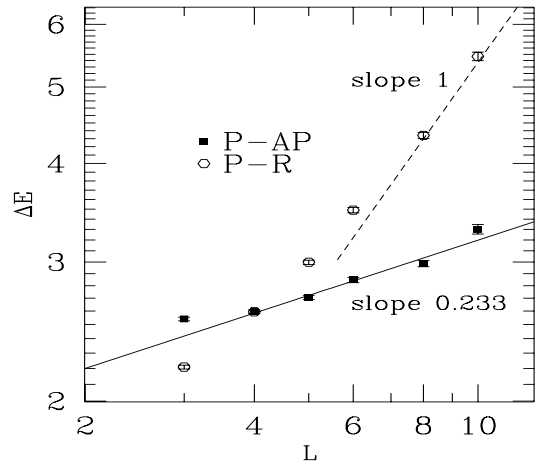,width=8cm,height=6.5cm}
\epsfig{figure=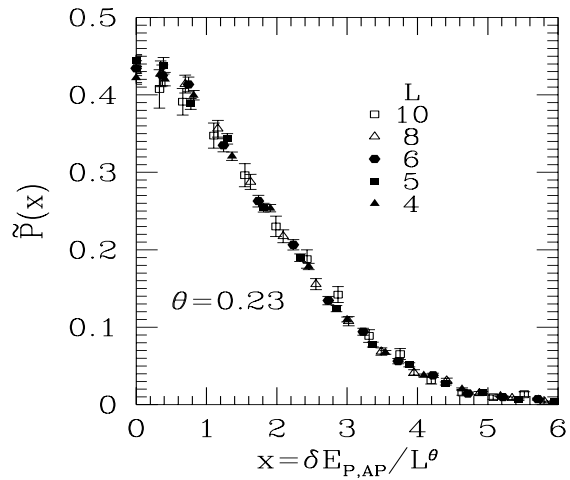,width=8cm,height=6.5cm}
\end{center}
\caption{
Upper figure: a plot of the root mean square ground state energy differences
$ \Delta E_{P,AP}$
and $\Delta E_{P,R}$ for different sizes up to $L=10$ in $d=3$. The fit for
the P-AP data omits the $L=3$ point.
Lower figure: a
scaling plot of the
symmetrized distribution of the values for $\delta E_{P,AP}$, the
energy difference between periodic and anti-periodic boundary conditions for a
single sample.  The distribution scales quite well with the value of $\theta$
obtained from the fit to $\Delta E_{P,AP}$.
}
\label{de2}
\end{figure}

Now we discuss our numerical results,
for which ake $L_B=2$. 
First of all, Fig.~\ref{de2} shows the root mean square energy difference,
$\Delta E$,
between P and AP and between P and R boundary conditions in $d=3$.
Apart from the
smallest size, $L=3$, the data for $\Delta E_{P,AP}$, are consistent with the
power law variation 
\begin{equation}
\Delta E_{P,AP} \sim L^\theta
\end{equation}
with 
\begin{equation}
\theta = 0.23^{+0.02}_{-0.04} ,
\label{theta_value}
\end{equation}
where the asymmetric error bar comes from systematic effects discussed below.
The positive value of $\theta$ shows that the system is stable against breaking
up into large domains of little energy, which implies that
$T_c > 0$, in agreement with earlier work\cite{ky,mari3d,hart,matteo}.
The value for $\theta$ is a little larger than earlier 
estimates\cite{other-theta} for the Gaussian distribution 
considered here, but these
calculations used a much smaller range of sizes.  For the $\pm J$
distribution, Hartmann\cite{hart} studied  sizes up to $L=10$ and found
$\theta = 0.19 \pm 0.02$, which is just consistent with the value here.
It is
expected\cite{fh} that
the distribution of energy differences, $\delta E_{P,AP}$ has the scaling form
$ P(\delta E_{P,AP}) = L^{-\theta} \widetilde{P}(\delta E_{P,AP}/L^\theta) $
and this works quite well as shown in the lower part of
Fig.~\ref{de2}.
As discussed in PY, $\Delta E_{P,R}$ 
is expected to 
vary as $L^{(d-1)/2},\ (= L$ here), for large $L$.
Fig.~\ref{de2} shows that our results are consistent with this behavior.

\begin{figure}
\epsfxsize=\columnwidth\epsfbox{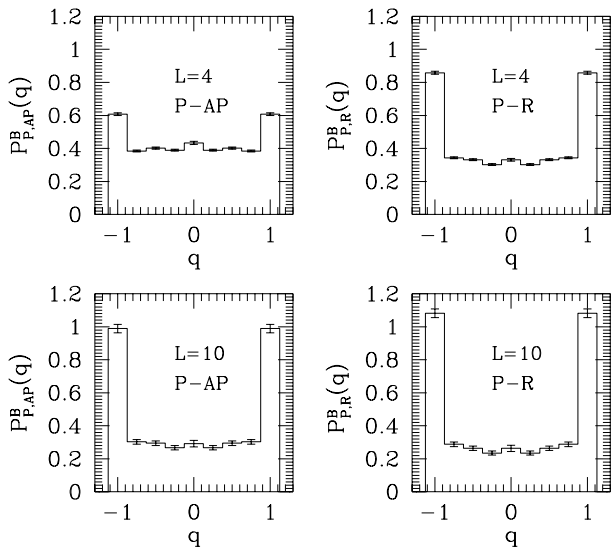}
\caption{
A plot of histograms of block spin overlaps for $L=4$
and 10 in $d=3$, with block size $L_B=2$.
Note that the allowed values of
$q$ are $0, \pm 0.25, \pm 0.5, \pm 0.75$ and $\pm 1$.
The left hand column is for the P-AP overlap and the right hand column for
the P-R overlap. The top row is for $L=4$ and the bottom row for $L=10$.  The
data is symmetrized and normalized so that the area under the histograms is unity.
}
\label{hist_multi}
\end{figure}

Some representative histograms of the block overlap distributions are shown in
Fig.~\ref{hist_multi}
for $d=3$. One sees that for both P-AP and P-R overlaps the weight
at $q=\pm 1$ increases with increasing $L$.

\begin{figure}[ht]
\begin{center}
\epsfig{figure=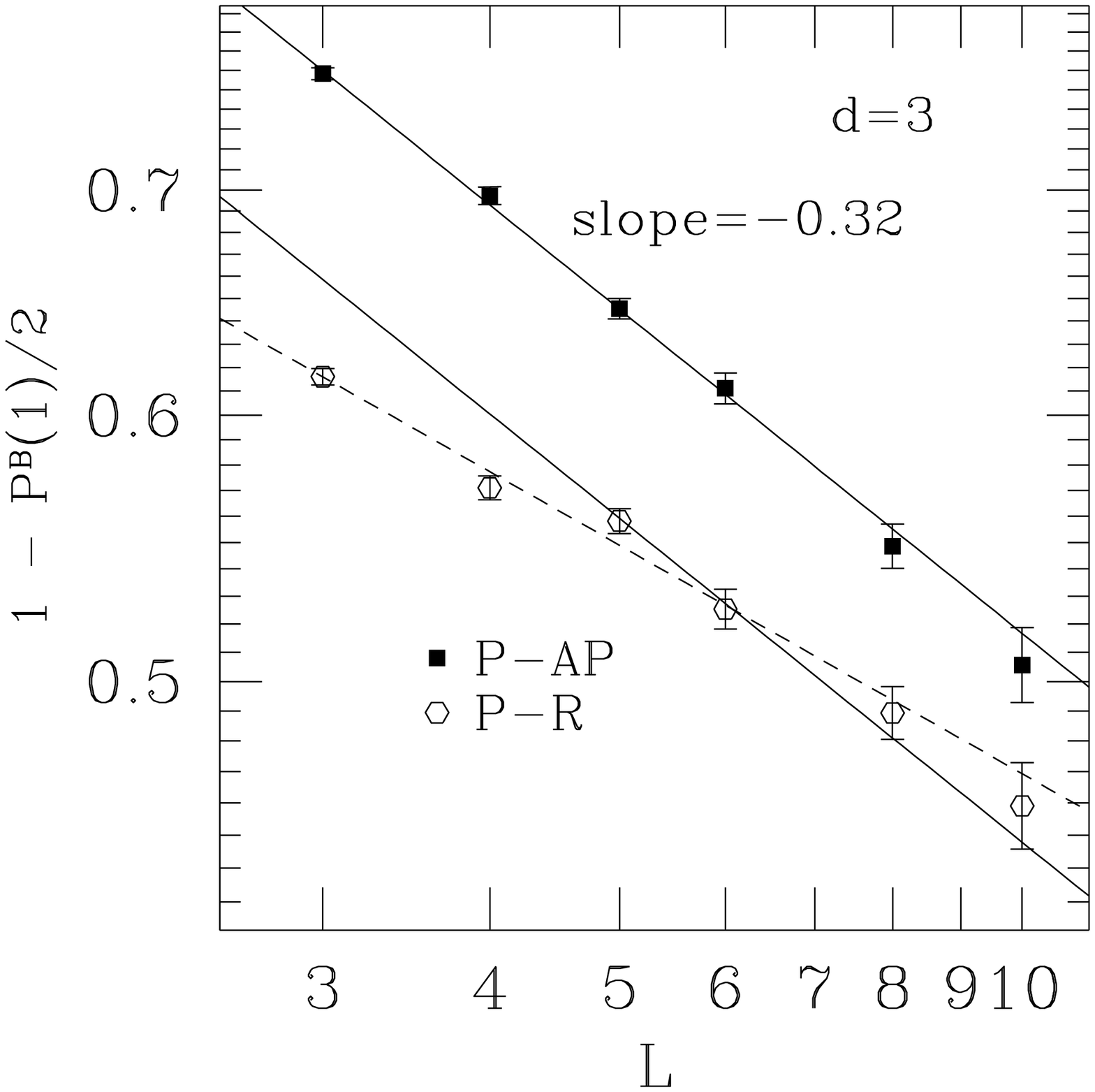,width=8.5cm,height=6.5cm}
\epsfig{figure=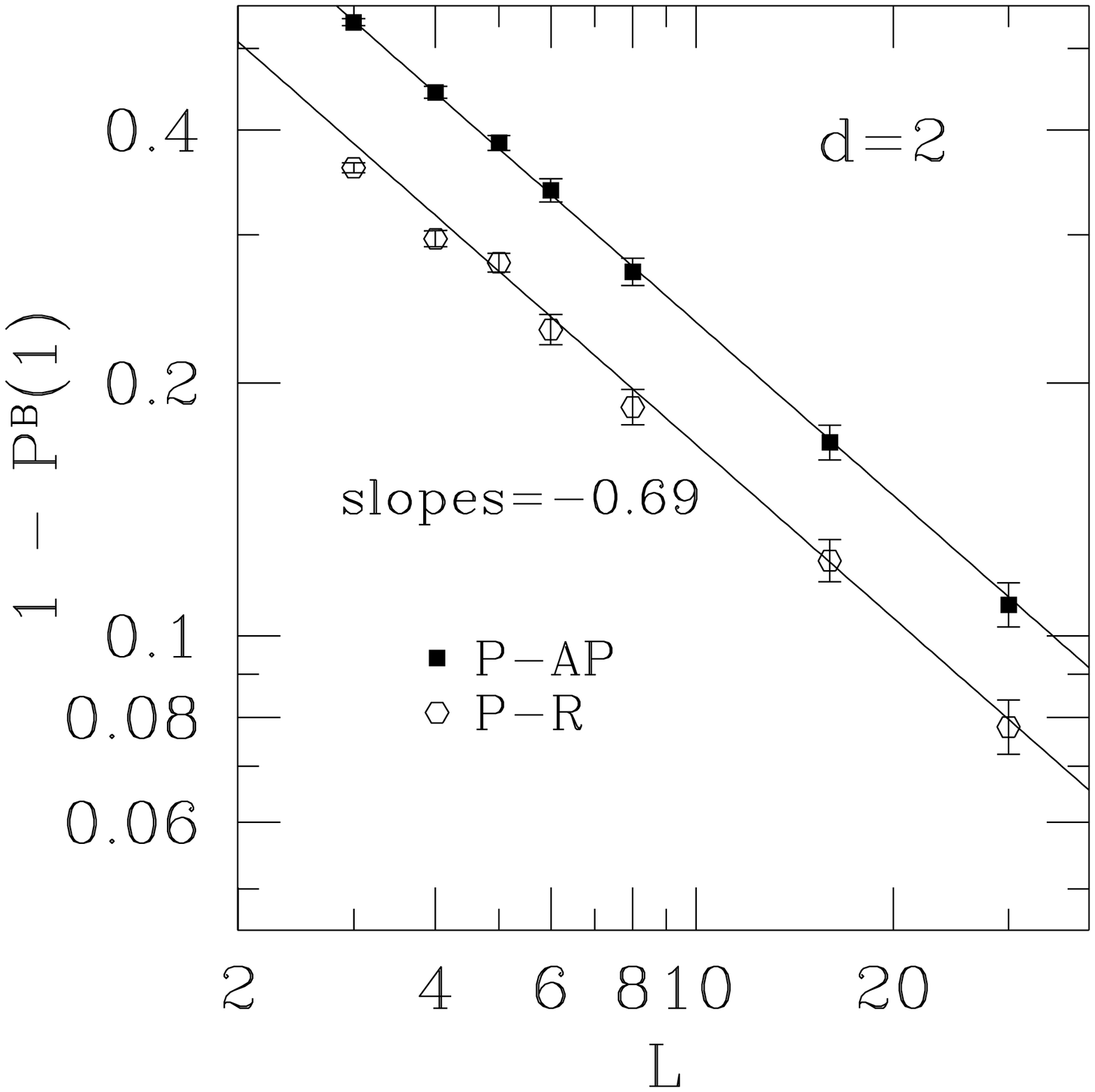,width=8.5cm,height=6.5cm}
\end{center}
\caption{
The figures show the probability that the spin configuration
of the central block 
changes when the boundary conditions are changed. 
Upper figure:
Data for three dimensions, for different sizes up to $L=10$
with block size $L_B$ equal to 2.
The solid line through the
P-R data is parallel to the fit for the P-AP points (slope $=-0.32$),
and fits the results for
large $L$. The dashed line is fit to {\em all} the P-R data and has a 
different slope
of $-0.23$.
Lower figure: a similar plot but for $d=2$, with
sizes up to $L=30$ and $L^B =2$.
}
\label{p1_3d}
\end{figure}

The probability that the configuration of
the block changes when the boundary conditions are changed, see
Eq.~(\ref{def_lambda}),
is plotted for a range
of sizes in the upper part of Fig.~\ref{p1_3d}.
For comparison, the lower part of the figure shows similar data
in $d=2$ (see PY for related results).

In $d=3$, the data for P-AP overlaps lies on a good straight line with a slope
$-\lambda$ where $\lambda  = 0.32 \pm 0.02$.
This leads to a fractal dimension of the domain
walls given by
\begin{equation}
d_f = d - \lambda = 2.68 \pm 0.02 
\end{equation}
The data for P-R overlaps for larger sizes lies parallel
to this, as shown in the figure,
but with some deviations for smaller sizes. However, given the statistical
uncertainties, one can also fit {\em all}
the P-R data to a power law
with a different
slope giving $\lambda = 0.23 \pm 0.02$.
If the ground state structure in $d=3$
were non-trivial the
data for the P-R boundary conditions in the upper part of Fig.~\ref{p1_3d}
would eventually saturate at a finite value for $L \to \infty$. There is no
sign of such a saturation for the sizes that we are able to study. We also
considered fits of the form $a + b L^{-\lambda}$, finding that, although a
range of positive values of $a$ is not ruled out by the data, the value $a=0$
is statistically preferred.

We also investigated the effects of
changing the bonds on one boundary to other random values
(rather than just changing the {\em sign})
and also changed the 
bonds on {\em all three} boundaries to new random
values. The results for these boundary conditions
give similar results to
those for the R boundary condition.

It is important to estimate the size of the errors which arise because the
algorithm is not guaranteed to find the exact ground state.
This is a problem only
for the larger sizes so we estimated the errors 
carefully for $L=8$ and $10$ for P-AP boundary conditions by
doing a smaller number of samples for a larger number of runs, $n_r=10$, than
before (see Table I) and assuming that the difference in results is a
reasonable measure of the error in the original data.
For $L=8$, the result for whether
the block spin configuration changed upon changing the boundary conditions was
the same after 2 runs as after 10 runs in 367 out of the 370 samples
considered, with 1
sample giving $|q|=1$ after 2 runs but $|q| < 1$ after 10 runs, and 2 samples
the other way round.
Hence our estimate of the relative error in the data in 
Fig.~\ref{p1_3d}
(due to not finding the correct ground state)
is in the range $(-1\pm \sqrt{5})/370$, i.e. 
between $-0.7 \%$ and $+0.2\%$, which is well within the 
statistical error bar of $\pm1.5\%$ shown in Fig.~\ref{p1_3d}.
For $L=10$ the corresponding figures are:
11 out of the 266 samples considered gave $|q|=1$ after 1 run but 
$|q| < 1$ after 10 runs, and 13
samples the other way round. The estimate of the error is therefore
$(-2 \pm \sqrt{24})/266$, i.e. the true answer should lie between
$-2.5\%$ and $+1.1\%$ of the value given, compared with the 
statistical error bar of $\pm 2.5\%$ in Fig.~\ref{p1_3d}. The error from 
the algorithm is therefore 
no bigger than the statistical error, and is significantly less than that on
the positive side. Hence the decreasing trend in the $3d$
data in Fig.~\ref{p1_3d}, is {\em not} due to
inaccuracies in the algorithm. Altogether, we find no evidence for
a {\em systematic} error in the data in Fig.~\ref{p1_3d} due to not always
finding the true ground state.
A similar analysis for the defect energy with P-AP boundary conditions
finds that, for $L=8$, the error 
is negligible compared with the statistical error, but that for $L=10$
the result is about $5\%$ too high, compared with the error bar in
Fig.~\ref{de2}
of about $\pm 1.5\%$. This is probably why the $L=10$ data point lies above the
fit. The best fit with the $L=10$
point $5\%$ lower has $\theta = 0.21$, rather than
0.23, which is why we give asymmetric error bars in Eq.~(\ref{theta_value}).

In $d=2$, the data for P-AP and P-R lie nicely parallel to each other
and we find $\lambda = 0.69 \pm 0.02$, and hence
$d_f = 1.31 \pm 0.02$, for both sets of boundary conditions,
in agreement with PY who
get $\lambda = 0.70 \pm 0.08$ and also in agreement with Middleton\cite{midd}.
Note that $\lambda$
is substantially smaller in $d=3$ than in $d=2$.
It would be interesting to know how it varies in higher dimensions,
but the numerics of such a calculation are challenging.

The only other calculation of $d_f$ in $d=3$
that we are aware of is that of Huse\cite{huse} who studied domain
growth at finite temperatures by Monte Carlo simulations, obtaining $d_f
\simeq 2.2$.  However, his method ``only measures the size of
domains which are {\em compact}\/'',
which may explain why his value is lower than ours.

To conclude we have seen that the ground state structure appears to be trivial
in a spin glass model with a finite $T_c$, the three-dimensional Ising spin
glass.  It remains to understand why
Monte Carlo simulations at finite 
temperature find,
by contrast,
evidence for a non-trivial pure state structure.

We would like to thank G. Parisi and D.~S.~Fisher for helpful comments on an
earlier version of this manuscript.
This work was supported by the National Science Foundation under grant DMR
9713977.  M.P. is supported in part by University of California, EAP Program,
and by a fellowship of Fondazione Angelo Della Riccia. The numerical
calculations were made possible by allocations of time from 
the National Partnership for 
Advanced Computational Infrastructures (NPACI) and
the INFM Parallel Computing Initiative.
We also thank Prof. M.~J\"unger and his group
for putting their spin glass ground state server in the
public domain.


\begin{references}
\bibitem{zero-field}
    For the case of zero field, which we consider here, there are actually two
    states, with the same free energy, related by flipping all the spins.
\end{references}
\end{document}